# Online detection of temporal communities in evolving networks by estrangement confinement

## Vikas Kawadia

vkawadia@bbn.com

Raytheon BBN Technologies, Cambridge MA 02138

## Sameet Sreenivasan

sreens@rpi.edu

Social and Cognitive Networks Academic Research Center,

Rensselaer Polytechnic Institute, Troy NY 12180

## Abstract

Temporal communities result from a consistent partitioning of nodes across multiple snapshots of an evolving complex network that can help uncover how dense clusters in a network emerge, combine, split and decay with time. Current methods for finding communities in a single snapshot are not straightforwardly generalizable to finding temporal communities since the quality functions used for finding static communities have highly degenerate landscapes, and the eventual partition chosen among the many partitions of similar quality is highly sensitive to small changes in the network. To reliably detect temporal communities we need not only to find a good community partition in a given snapshot but also ensure that it bears some similarity to the partition(s) found in immediately preceding snapshots. We present a new measure of partition distance called *estrangement* motivated by the inertia of inter-node relationships which, when incorporated into the measurement of partition quality, facilitates the detection of meaningful temporal communities. Specifically, we propose the estrangement confinement method, which postulates that neighboring nodes in a community prefer to continue to share community affiliation as the network evolves. Constraining estrangement enables us to find meaningful temporal communities at various degrees of temporal smoothness in diverse real-world datasets. Specifically, we study the evolution of voting behavior of senators in the United States Congress, the evolution of proximity in human mobility datasets, and the detection of evolving communities in synthetic networks that are otherwise hard to find. Estrangement confinement thus provides a principled approach to uncovering temporal communities in evolving networks.

# 1   Introduction

Community detection has been shown to reveal latent yet meaningful structure in networks such as groups in online and contact-based social networks, functional modules in protein-protein interaction networks, groups of customers with similar interests at online retailers, disciplinary groups of scientists in collaboration networks, etc.[1]. Temporal community detection aims to find how such communities emerge, grow, combine and decay in networks that evolve with time. Temporal communities can be used to detect trends, anomalies and other events of interest in complex time evolving networks. They can provide robust network-based insights into the evolution of inter-country trade networks, the emergence of celebrities in social media, the growth and division of distinct political ideologies in congress etc. Temporal communities also have direct applications like informing public health policy in countering epidemics, and identifying business opportunities in the technology sector.

Static community detection [1] aims to partition a network into groups of nodes such that the intra-group edge density is higher than the inter-group edge density. A partition $P$ can be specified by labels $\{l_1, l_2, ..., l_N\}$ assigned to the $N$ nodes in the network. Each group (nodes with the same label) in the partition then constitutes a community. If the network is time-varying, then given snapshots $\{G_t\}$ ($0 \leq t \leq T$), temporal community detection assigns labels to nodes in each snapshot, and the set of {node, time} pairs that get the same label constitute a *temporal community*. That is, a temporal community structure is a partitioning of the {node, time} pairs in all the snapshots that optimizes an appropriate quality function. Mucha et al.[2] proposed one such quality function - multi-slice modularity - that is defined on a stacked aggregate network consisting of all snapshots. In this work, we focus on the *online version* of the temporal community detection problem, where one is allowed to do computations only on the current snapshot while using limited information from the past. This is useful in situations where the number of snapshots is large, or fast computation of temporal communities is important as new snapshots become available.

A popular approach to detecting temporal communities is to find static communities independently in each snapshot using some quality function and then "map" communities between snapshots to preserve labels when possible. For example, Rosvall et al.[3] use a significance based method to match communities obtained from the map-equation in successive snapshots. Palla et al.[4] use clique percolation to find communities within individual snapshots and map them by finding clique communities in a merged network of adjacent snapshots. Greene et al.[5] define temporal communities as moving *fronts* and propose a rule based on the Jaccard overlap to map communities in the current snapshot to existing fronts. However, none of the above methods explicitly use the knowledge of partitions found in past snapshots to inform the search for the optimal partition on the current snapshot. We argue (and show empirically in *Results*) that they are likely to miss crucial temporal communities because of the challenges inherent even in static community detection. Methods used to discover communities in static networks are based on finding a partition of nodes that optimizes some quality (objective) function that quantifies how community-like the partition is. One of the earliest proposed and still commonly used quality functions is *modularity* [6], although several others have been subsequently proposed. Good et al.[7] have shown that for many real-world networks, the modularity landscape is highly degenerate and disordered with numerous partitions that differ only slightly, yielding similar values of modularity and constituting distinct local maxima of the landscape. Importantly, as pointed out in [7], similar issues of degeneracy are also likely to be present for various other quality functions. Moreover, the quality function landscape could be highly sensitive to changes in the network, making it very likely that a rather distinct community partition is detected even when only a few nodes and edges are added or deleted to the network. For example, Karrer et al. [8] have demonstrated such a sensitivity for the modularity landscape of several synthetic and real-world networks. This degeneracy of the quality function landscape and its sensitivity to small changes is what primarily makes mapping independently detected communities across snapshots difficult.

We propose a method that aims to find a good partition in a given network snapshot by exploiting the knowledge of the community structure in the previous snapshot. The rationale behind the method is that any system tends to maintain some temporal contiguity in its features as it evolves. Obviously, independent maximization of modularity (or some other quality function) on each snapshot has no incentive to maintain such a temporal contiguity between partitions. To find meaningful temporal communities, it is crucial to narrow the search to those partitions in the current snapshot that bear some similarity to the partitions found in the previous snapshots. One of the key challenges is to find a measure of this partition-similarity (or distance) that is appropriate for comparing partitions of different snapshots of an evolving network. None of the existing measures of partition distance, such as Variation of Information [7], are suitable for comparing partitions of nodes in distinct snapshots across which the network can change. This is because they do not consider edges in the network, and therefore cannot account for changes in network structure. In particular, we require a measure that ignores differences in node partitions when the network has changed significantly (between the two snapshots we are comparing), but penalizes (i.e. yields a large distance for) dissimilar partitioning when there are only minor changes in network.

We present a novel measure of partition distance, called *estrangement*, which quantifies the extent to which neighbors continue to share community affiliation. This is motivated by the empirical observation that it is some form of social inertia inherent to group affiliation choices that prevents the community structure from changing abruptly [9, 10]. The estrangement between two time-ordered snapshots is defined as the fraction of edges that stop sharing their community affiliation with time, as illustrated in Fig. 1. Our method of detecting temporal communities consists of maximizing modularity in a snapshot subject to a constraint on the estrangement from the partition in the previous snapshot. The amount of estrangement allowed controls the *smoothness* of the evolution of temporal communities, and varying it reveals various levels of resolution of temporal evolution of the network. The estrangement constrained modularity maximization problem described above is at least as hard as modularity maximization which is NP-complete [11]. Moreover, known heuristic methods for unconstrained modularity maximization are not directly applicable to the constrained version. However, we show that the dual problem constructed using Lagrangian relaxation can be tackled by adapting techniques used for unconstrained modularity maxi-mization, specifically a version of the Label Propagation Algorithm [12, 13].

Some recent proposals for detecting temporal communities, similarly to ours, use the past community structure to improve community coherence. Mucha et al. [2] extend the notion of random walk stability, first introduced by Lambiotte et al. [14], to mutli-slice networks and show that optimization of this stability yields coherent temporal communities. However, their method requires all slices (snapshots) to be aggregated into a stacked graph by introducing arbitrary weighted links between node copies in different slices, and is thus not amenable to the online temporal community detection problem, in contrast to the method we propose. (Incidentally, estrangement can be interpreted as *temporal stability* as shown in the SI. Our method is closest to *evolutionary clustering* introduced by Chakrabarti et al. [15] where the quality of a community partition is measured by a combination of its *snapshot cost* and its *temporal cost*. However, unlike our method, [15] does not prescribe specific relative contributions of the two costs, or demonstrate the effect of varying these contributions. Furthermore, the partition distance measure and the optimization techniques we use are different from those in [15]. Subsequent techniques such as *FacetNet* [16] and *MetaFac* [17], apply the evolutionary clustering approach to partitions derived from a generative mixture model approximation of the network adjacency matrix. Broadly applicable generative models are, however, challenging to create for real complex networks. GraphScope [18] finds temporal communities by breaking the sequence of graph snapshots into *graph segments* and finding good communities within each graph segment such that the total cost of encoding the sequence of graphs is minimized. However, it can only be used on unweighted networks. The method we propose works in an online setting, does not need any generative model for network structure or evolution, and is applicable to both weighted and unweighted networks. Next, we present a precise formulation for the temporal community detection problem and describe our solution method.

## 2 Methods

### 2.1 Problem Formulation

Given network snapshots $G_{t-1}, G_t$ and the partition $P_{t-1}$ that represents the community structure at time $t-1$, find a partition $P_t$ of $G_t$ that solves the following constrained optimization problem:

$$\underset{\mathcal{P}}{\text{maximize}} \quad Q(\mathcal{P})$$
$$\text{subject to} \quad E(\mathcal{P}) \leq \delta. \tag{1}$$

Here $Q$ is a quality function for the community structure in a snapshot, $\mathcal{P}$ denotes the space of all partitions, and $E$ is a measure of distance or dissimilarity between the community structure at times $t$ and $t-1$. The formulation above is based on the intuition that temporal communities can be detected by optimizing for quality in the current snapshot while ensuring that the distance from the past community structure is limited to a certain amount, as specified by the parameter $\delta$. Smaller values of $\delta$ imply greater emphasis on temporal contiguity whereas larger values of $\delta$ place greater focus on finding better instantaneous community structure. Hence, we refer to $\delta$ as the *temporal divergence*, or simply divergence. We emphasize that our formulation is independent of the specific community structure quality function used. In this paper, we use modularity [6], a widely studied and tested quality function, which is defined as follows:

$$Q = \frac{1}{2M} \sum_{u,v} (A_{uv} - \frac{k_u k_v}{2M}) \delta(l_u, l_v), \tag{2}$$

where $A$ is the adjacency matrix for the network, $k_x$ is the degree of node $x$ and $l_x$ is the label assigned to $x$ in this partition and $M$ is the total number of edges in the network. $\delta(i,j)$ is 1 if and only if $i = j$, and 0 otherwise. Modularity scores a partition by the fraction of intra-community edges in the network less the same fraction in an appropriate random rewiring of the network.

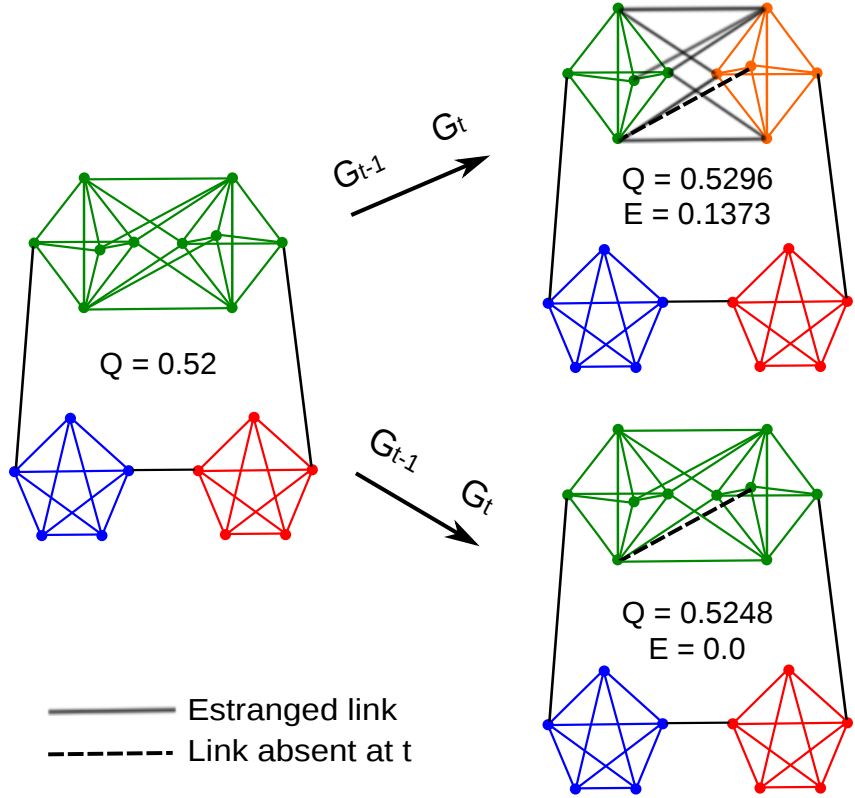

Figure 1: An example illustrating the detection of temporal communities via estrangement confinement. The network on the left, $G_{t-1}$ consists of 20 nodes and 52 links, and a maximal modularity partition of this network consists of three communities represented by the three colors ($Q = 0.52$). In the next snapshot, the network has evolved to $G_t$ which differs from $G_{t-1}$ only in the absence of a single link, indicated by the dotted line. The top right and bottom right networks both represent the same network $G_t$, but indicate distinct choices of community partitions available. The partition shown on the top right, $P_1^t$ consists of 4 communities, and is the partition that gives the highest modularity $Q_1^t = 0.5296$. The partition $P_2^t$ for $G_t$ shown on the bottom right which preserves the node partition chosen for $G_{t-1}$ has a slightly lower modularity of $Q_2^t = 0.5248$. The partition $P_1^t$ with higher modularity, however, makes 7 links *estranged*. The estranged links (shown in gray) are those links present at both $t-1$ and $t$ and whose endpoint nodes shared community affiliation at $t-1$ but no longer do so at $t$. Notice that links in the orange community of $P_1^t$ despite having changed their community affiliation from $t-1$ to $t$ are not estranged since their endpoint nodes still share community affiliation. In contrast to $P_1^t$, the partition $P_2^t$ yields no estranged links. Estrangement, $E$, defined as the fraction of estranged links at $t$ is therefore zero for $P_2^t$ but $7/51 = 0.13$ for $P_1^t$. Maximizing modularity while constraining estrangement to a low value (like 0.01) therefore chooses $P_2^t$ as the partition for $G_t$, yielding a smoother temporal progression of the community structure from $t-1$ to $t$.

For measuring partition distance, we use our novel measure of estrangement which we now define precisely. Given network snapshots $G_{t-1}, G_t$ and partitions $P_{t-1}$ and $P_t$, an edge $(u, v)$ in $G_t$ is said to be *estranged* if $l_u \neq l_v$ in $P_t$, given that $u$ and $v$ were neighbors in $G_{t-1}$ and $l_u = l_v$ in $P_{t-1}$. Estrangement is now defined as the fraction of estranged edges in $G_t$. Note that equality of labels is required only within partitions, not across partitions. Estrangement can be written as:

$$E = \frac{\sum_{u,v \in G_t} Z_{uv}(1 - \delta(l_u^t, l_v^t))}{2M} \tag{3}$$

where $Z_{uv} = \delta(l_u^{t-1}, l_v^{t-1})\sqrt{A_{uv}^{t-1} A_{uv}^t}$, where $A^{t-1}$ and $A^t$ are the adjacency matrices of $G_{t-1}$ and $G_t$ respectively. The square root term ensures that the definition applies to weighted networks as well (see SI). Estrangement can take values between 0 and 1, with 0 estrangement implying maximum possible similarity between the community structure in the two snapshots of the network and a value of 1 implying maximum possible dissimilarity.

## 2.2 Duality based optimization approach

Greedy local optimization methods used for modularity maximization cannot be directly used to solve the constrained optimization problem in Eq. 1, since the space of solutions is now confined to the set of partitions which respect the constraint. We use the Lagrangian duality approach for constrained optimization. This first needs computing the Lagrange dual function, which we show can be computed by adapting known methods for unconstrained modularity maximization. The key to computing the dual lies in exploiting the property that estrangement is decomposable, similarly to modularity, into single node (or local) contributions. We adapt a hierarchical version of the Label Propagation Algorithm [12] to compute the dual. This method works by greedily merging communities that provide the largest gain in the objective function, and then repeating the procedure on an *induced* graph in which the communities from the previous steps are the nodes. Once this method of computing the Lagrange dual has been determined, we solve the dual problem of finding the best Lagrange multiplier by using Brent's method which is commonly used for non-differentiable objective functions. We now present the above steps in greater detail.

Henceforth, for notational simplicity, unless otherwise stated, all quantities of interest are with respect to the current snapshot $G_t$. Following the dual formulation [19], we write the Lagrangian $L$ and the Lagrange dual function $g$ corresponding to the primal problem (Eq. 1) as:

$$
\begin{aligned}
L(\mathcal{P}, \lambda) &= Q - \lambda \, (E - \delta) \\
g(\lambda) &= \sup_{\mathcal{P}} L(\mathcal{P}, \lambda)
\end{aligned}
\tag{4}
$$

where $\lambda$ is the Lagrange multiplier. For every value of $\lambda$, the function $g(\lambda)$ yields an upper bound to the optimal value $Q^*$ of the primal problem. We are interested in the value of $\lambda$ that yields the smallest upper bound, which would in turn give us the best estimate of $Q^*$ subject to the constraint on $E$. This dual problem corresponding to the primal problem in Eq. 1 is as follows:

$$
\begin{aligned}
&\text{minimize} \quad g(\lambda) \\
&\text{subject to} \quad \lambda > 0
\end{aligned}
\tag{5}
$$

If the minimum of $g(\lambda)$ occurs at $\lambda^*$, the optimal partition for a given snapshot is one that yields the supremum of $L(\mathcal{P}, \lambda^*)$ over all partitions. To find the best partition, we need to solve the dual problem which needs computing the Lagrange dual function. Both involve challenges that we discuss below.

## 2.3 Computing the Lagrange dual

To compute $g(\lambda)$, we need to search the space of partitions for one that yields the supremum of $L(\mathcal{P}, \lambda)$. We show that this is no harder than unconstrained modularity maximization, and known methods for $Q$ maximization can be adapted. We use a hierarchical adaptation of the Label Propagation Algorithm (LPA), that we call HLPA, to do a multi-level local greedy search in partition space. LPA [13] is an agglomerative approach to finding communities, where each node is initialized with a unique label and at each subsequent step of the algorithm, a node updates its label to the current most common label in its neighborhood. In general, variants of LPA tailored to specific optimization problems can be constructed by modifying the local objective function that the label update is maximizing. Barber and Clark [12] proposed one such variant for modularity maximization that they referred to as LPAm. We construct the label update rule in HLPA in a similar vein, with the optimization given by Eq. 4 in mind. The reason this label propagation approach works for optimizing $L(\mathcal{P}, \lambda)$ is because estrangement can be decomposed into node-local terms which allows $L$ to be optimized by each node updating its label based on those in its neighborhood. Recall that a partition $P$ is specified by the labels $\{l_1, l_2, ..., l_N\}$ assigned to the nodes. Then, in a label update in HLPA, each node $x$ updates its community identifier $l_x$ following the rule:

$$
l_x = \arg\max_l \left( N_{xl} - \frac{k_x K_l}{2M} + \frac{k_x^2}{2M} \delta(l_x, l) + \lambda O_{xl} \right),
\tag{6}
$$

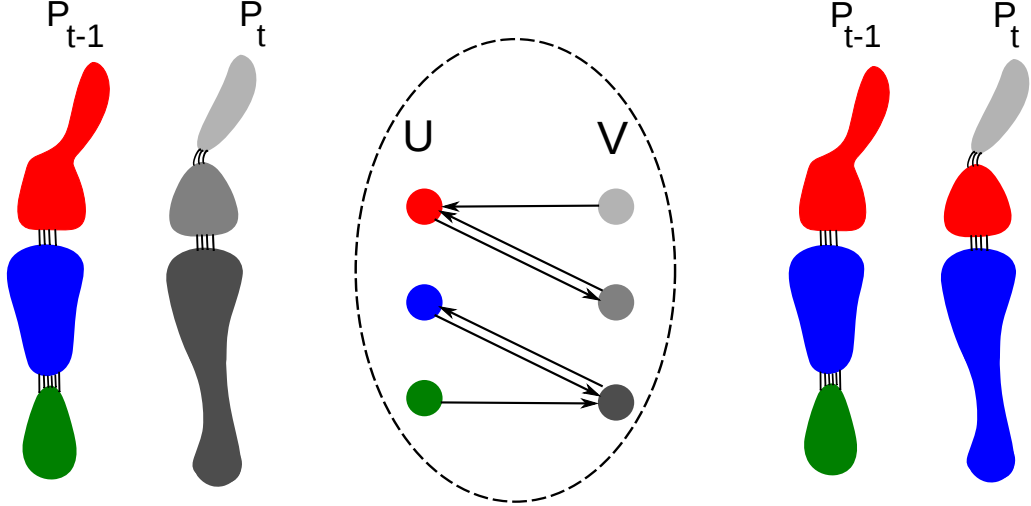

Figure 2: Mapping community labels from time $t - 1$ to time $t$. The left panel shows the situation after estrangement confinement has found a partition of the graph at time $t$, consisting of 3 communities. Two of these have arisen due to an uneven split of the red community at $t - 1$, and one due to the merging of the blue and green communities at time $t - 1$. The intent of the mapping procedure is to cause fewest nodes to change labels from $t - 1$ to $t$. The center panel shows the bipartite construction that the mapping procedure relies on. Here, nodes on the left (set $U$) represent communities at $t - 1$ and nodes on the right (set $V$) represent those at time $t$. Each node in $U$ has an outgoing link to the node in $V$ with whom its Jaccard overlap is maximal. Similarly each node in $V$ has an outgoing link to the node in $U$ with whom its Jaccard overlap is maximal. For simplicity, we say that each node points to its maximal overlap partner in the other set. Once these links are drawn, the mapping procedure allows inheritance of labels only between pairs of nodes which have bidirectional links between them, i.e., a node in $U$ (community at $t - 1$) passes on its label to a node in $V$ (community at $t$) only if they are maximal overlap partners of each other. Consequently, a node in $U$ whose outgoing link and incoming link do not share the same partner node in $V$, does not pass on its label (For example, the green node in $U$). Similarly, a node in $V$ whose outgoing link and incoming link do not share the same partner node in $U$, does not inherit any label, and therefore obtains a new label (for example, the topmost node in $V$). The progression of appropriately labeled communities from $t - 1$ to $t$ after the mapping step is shown in the panel on the right.

where $N_{xl} = \sum_u A_{ux}\delta(l_u, l)$, $O_{xl} = \sum_{u \neq x} Z_{ux}\delta(l_u, l)$ and $K_l = \sum_u k_u\delta(l_u, l)$. Here $O_{xl}$ is the extra term that arises due to the constraint on $E$. We show in the SI that the above update rule converges to a local optimum of $L(\mathcal{P}, \lambda)$.

The optimization of $L$ in HLPA is further improved by the following hierarchical procedure. If a sequence of label updates has converged on the original graph on which $L$ is being maximized, build a new *induced graph* which contains the communities of the original graph as nodes, and links between pairs of nodes in the new graph with weight equal to the total number of links between the two communities in the original graph. As we show in the SI, $L$ is invariant under this transformation. Nodes on the induced graph also iteratively update their labels following Eq. 6, until the labels have converged. This alternating procedure of label updates followed by the induce graph transformation is recursively applied until we reach a hierarchical level where the converged value of $L$ is lower than that obtained at the previous level. The partition found at the penultimate level before termination is chosen as the one optimizing $L$. This hierarchical procedure for optimizing $L$ is similar in spirit to the one used in the Louvain algorithm [20] for optimizing $Q$.

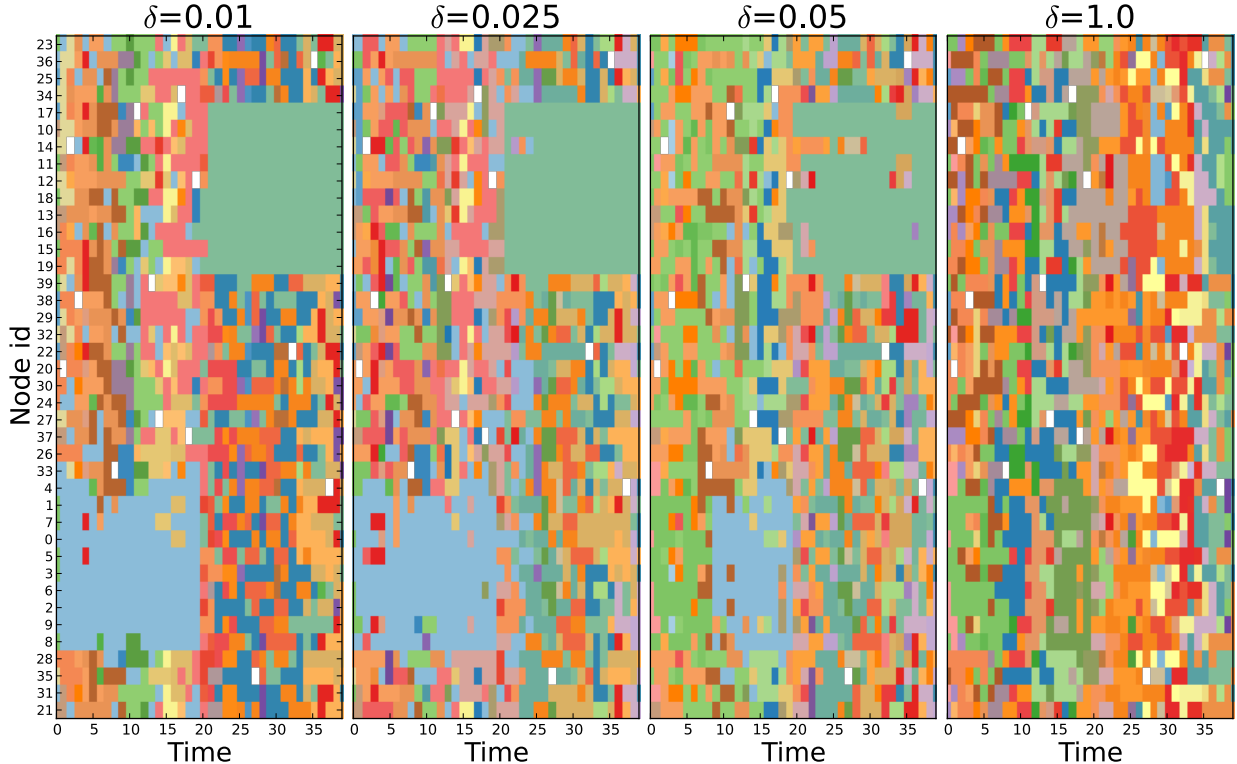

Figure 3: Estrangement confined modularity maximization allows detection of densely inter-connected groups embedded within a random background (see text for details) even when the groups randomly reshuffle their links. The ability to detect these *hidden* groups diminishes as the constraint on estrangement is relaxed ($\delta \to 1$) with the poorest results obtained for independent modularity maximization ($\delta = 1$).

## 2.4 Solving the dual problem

Having found a way to compute $g(\lambda)$ we can solve the dual problem and determine the value of $\lambda$ at which $g(\lambda)$ is minimized. The challenge here is that $g(\lambda)$ is not differentiable and moreover expensive to evaluate. We use Brent's method [21], which is often used to optimize non-differentiable scalar functions within a given interval. Furthermore, to mitigate issues due to the local nature of the algorithm and the degeneracy of the modularity landscape (and therefore the $L(\mathcal{P}, \lambda)$ landscape), we perform an adaptive number of several independent runs of HLPA for a given $\lambda$ and pick the run which yields the highest value of $g(\lambda)$. Please see the SI for details.

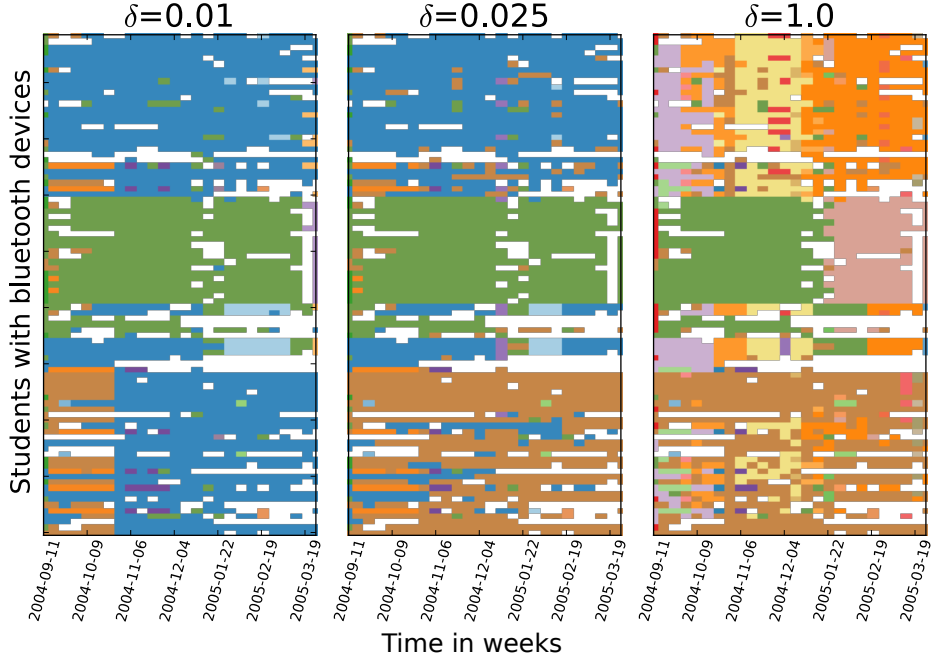

Figure 4: Temporal communities seen in the reality mining network. The network consists of two communities predominantly at lower values of $\delta$, one corresponding to staff and students at the MIT Media Lab, and the other corresponding to students at the MIT Sloan School of Business. As $\delta$ is increased the average size of the communities decreases and the number of communities increases, a consequence of the decreasing temporal contiguity.

## 2.5 Mapping communities

Tracking the evolution of communities requires the communities at time $t$ to be mapped to those at $t-1$. We map those communities across two consecutive snapshots that have the maximal mutual Jaccard overlap between their constituent node-sets (Jaccard similarity or overlap of two sets is defined as size of their intersection set divided by the size of their union), and generate new identifiers when needed (see Fig. 2). Other methods such as alluvial diagrams [3], and threshold based mapping [5], have been proposed for mapping community identifiers. Our method causes the fewest number of nodes that are common to both snapshots to change labels from one snapshot to the next.

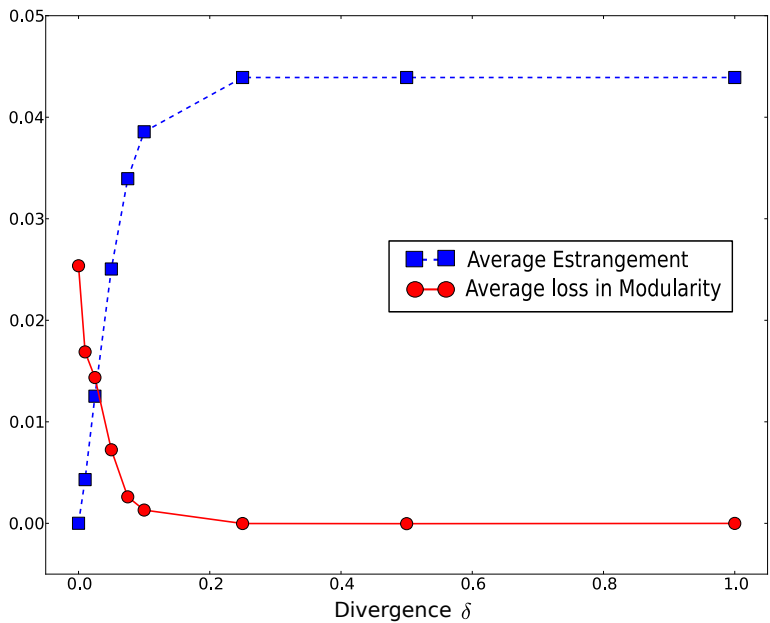

Figure 5: A tighter constraint on estrangement (lower $\delta$) implies a greater average loss in modularity.

## 3 Results

We apply the estrangement confinement method on one synthetic and two empirical networks, and use an impressionistic visualization to show the temporal communities that we refer to as an *evolution chart*, where nodes are shown along the Y axis and the snapshot number along the X axis. See Fig. 3 for an example. Each "pixel" in the evolution chart corresponds to a particular node at a given time, and the color represents the label of that node at that time $t$. The nodes are ordered on the Y axis by the tuple of labels they take over time, where the labels in the tuple itself are ordered by the frequency of acquiring that label. Ties are broken by the time of first appearance of nodes. This ordering causes the nodes in a temporal community to appear contiguously.

First, using a 40 node synthetic time-varying network, we demonstrate that our method can detect hidden node groups in networks that randomly vary their intra-group links to avoid detection. Each of the snapshots of the synthetic evolving network is generated from a 40 node Erdős-Rényi random network with 80 edges by taking a subset of nodes (nodes $0-9$ in the first twenty snapshots, and nodes $9-19$ in the last twenty snapshots) and adding 20 more edges between their members randomly. Thus the overall picture is that of an evolving random network containing within it two groups with higher intra-group link density, active at different periods in time, and whose intra-group connections are in constant flux. We observe in Fig. 3 that independent Q maximization (corresponding to $\delta = 1$) followed by mapping of labels is unable to detect the temporal communities of interest. However as $\delta$ is decreased, the temporal continuity constraint becomes stronger and improves the detection of the two hidden cores.

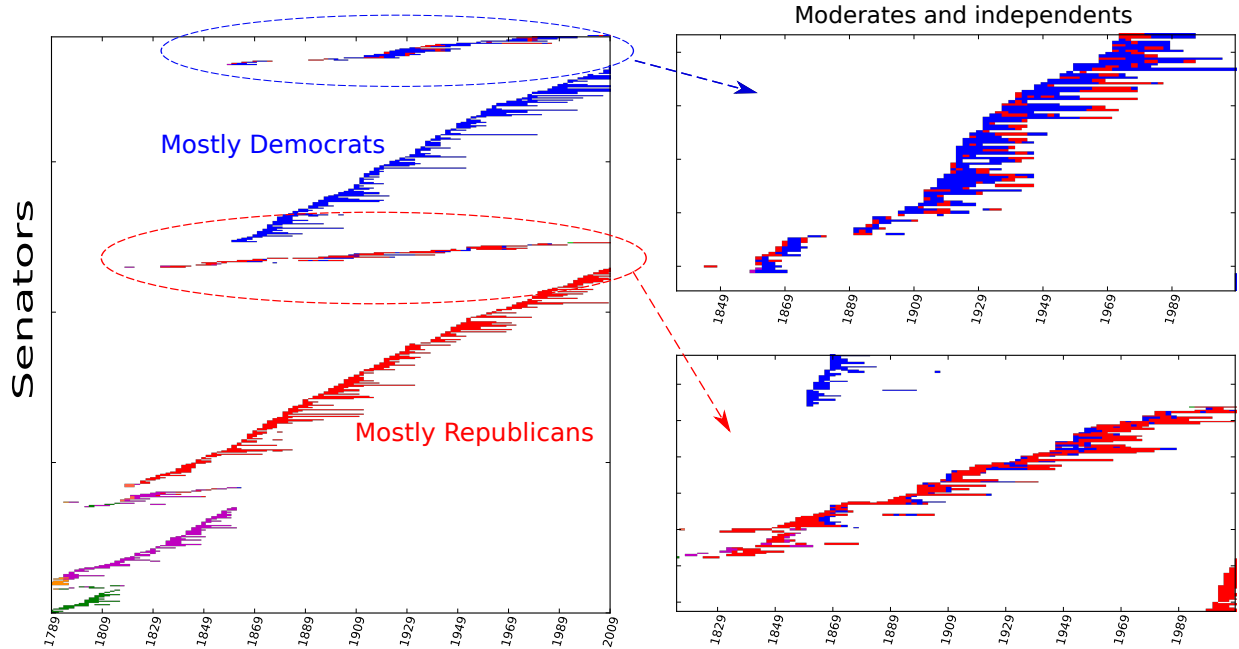

Figure 6: The different temporal communities observed in the senator voting similarity network. See text for details.

The effect of varying $\delta$ is also visible in our analysis of a real network: the human-human contact network data provided by the Reality-mining project [22] which tracked the mobility of about hundred individuals over nine months. A contact is registered when the Bluetooth devices being carried by the individuals come within 10m of each other. The evolution chart in Fig. 4, shows the temporal communities resulting from applying estrangement confinement to snapshots created by aggregating contacts over a week (except over vacation weeks in December) between individuals thus creating a weighted time evolving network, where in each snapshot the weight on an edge represents the number of contacts between the corresponding individuals. We illustrate communities and events that can be correlated with ground truth in Fig. 4.

Finally, we analyze a time-evolving weighted network consisting of United States senators where the weight on an edge represents the similarity of their roll-call voting behavior. The data was obtained from *voteview.com* and the similarities between a pair of senators was computed following Waugh et al. [23] as the number of bills on which they voted similarly, normalized by the number of bills they both voted on. The network consists of 111 snapshots corresponding to congresses over 220 years and 1916 unique senators. Although sweeping through different values of $\delta$ provides insights at different temporal resolutions, we can find a suitable value of $\delta$ by investigating the trade-off between estrangement and modularity as $\delta$ is varied. Fig. 5 shows the average loss (over all snapshots) in $Q$ and the average $E$ as $\delta$ is varied for the senator-voting dataset. For this dataset, we focus on the evolution chart for $\delta = 0.05$, the value at which the two curves in Fig. 5 intersect, representing a natural constraint to seek on estrangement. A broad feature that is observed for all values of divergence is the emergence of two dominant voting communities with time. The party affiliation of the majority of the constituent nodes within these communities allows us to identify them as the temporal streams which culminate in the present day Democratic and Republican parties. These features were previously observed by Mucha et al.[2]. We reiterate however, that in contrast to their method, ours is online and does not need to analyze the stacked network comprising all snapshots. In addition to the dominant Democratic and Republican streams, we also observe two minor communities that consist of senators who predominantly vote in alignment with one of the two dominant communities, but have occasional switches to the other. One of these communities – the central one – consists predominantly of Democratic members of the *conservative coalition.* The second such community – the topmost one – consists of several moderate Democrats and left-leaning Republicans (see inset). Another feature we find is the reduction with time in the number of senators whose aggregating voting behavior over the duration of a congress are not aligned with the rest of their party. Fig. 7 shows the number of such "atypical" senators over time. Notice that after the year 1995, there is only one such senator detected by our method, whereas prior to 1995, a much larger number of senators voted differently from the bulk of their party.

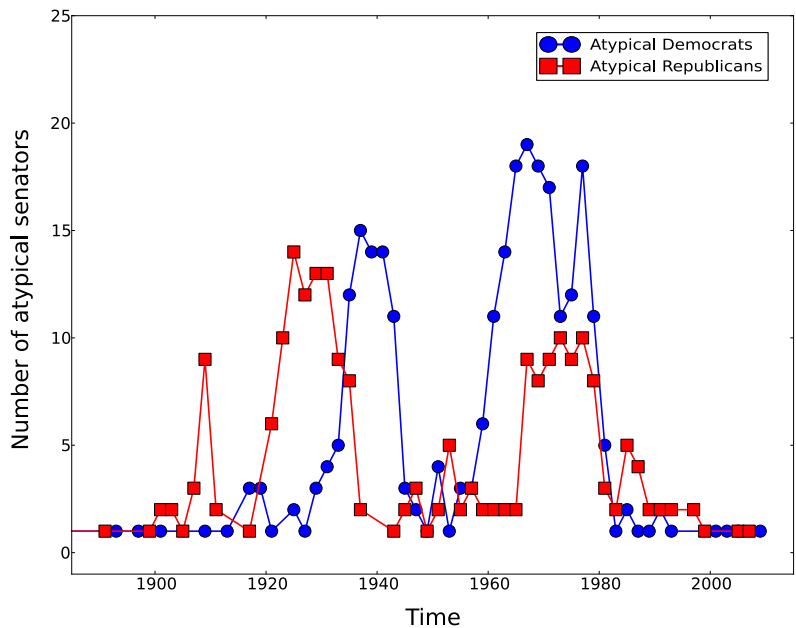

Figure 7: Extent of atypical voting in the US senate as inferred from temporal communities.

## 3.1 Discussion

We have presented a constrained optimization framework to find temporal communities and have introduced estrangement, an effective measure of partition distance between snapshots of a time-varying network. We demonstrate that meaningful temporal communities can be found by estrangement constrained modularity maximization, a problem we solve using Lagrange duality. Specifically, the fact that estrangement can be decomposed into local, single node terms enables efficient solution of the Lagrange dual problem through agglomerative greedy search methods.

Several outstanding issues are worthy of further study. An important one is of determining the granularity at which the time varying network is snapshotted. If the snapshots are made too frequently, there may not be enough density of edges to discover communities, while aggregating for too much time may prevent detection of some evolving patterns. In this work, we assume that there is a natural timescale of interest for creating snapshots, such as the one defined by biennial congressional elections in the case of the senator voting similarity network. In general, such natural timescales can perhaps be found by analyzing the frequency spectrum of some relevant variable in the dataset [22]. A related issue is that of sporadic interruptions in data collection which could affect the calculation of estrangement as well as the mapping between communities in successive snapshots. The effect of interruptions can perhaps be mitigated by maintaining a history of the extent to which two nodes share community affiliation and using it to compute estrangement. Secondly, estrangement appears to be easily generalizable to temporally evolving overlapping communities (see SI) which could reveal further interesting features in community evolution. Also, generalization of our method to directed networks remains to be investigated. Finally, one could investigate the effect of confining higher order estrangement terms rather than the dyadic form that we have used.

**ACKNOWLEDGEMENTS:** Research was sponsored by the Army Research Laboratory and was accomplished under Cooperative Agreement Number W911NF-09-2-0053. The views and conclusions contained in this document are those of the authors and should not be interpreted as representing the official policies, either expressed or implied, of the Army Research Laboratory or the U.S. Government. The U.S. Government is authorized to reproduce and distribute reprints for Government purposes notwithstanding any copyright notation here on.

# 4 Supplementary Information

## 4.1 Estrangement for weighted networks

Estrangement defined in Eq. 3 applies to to weighted networks as well by considering $M$ to be sum of the weights of all the edges in the network. The $\sqrt{A_{uv}^{t-1}A_{uv}^t}$ term implies that if the weight of an edge whose endpoints continue to share labels changes from time $t-1$ to $t$, we take the geometric mean of the weights when computing the partition distance. Modularity (Eq. 2) and our HLPA optimization technique (Eq. 6), can also be generalized to weighted networks by considering $k_u$ to be the *strength* of node $u$ instead of the degree, where strength is defined as the sum of the weights of adjacent edges, and by considering $M$ to be sum of the weights of all the edges in network. We have applied estrangement confinement to weighted networks in the results that we present in the main paper.

## 4.2 Estrangement and overlapping communities

Communities in networks often overlap such that nodes can simultaneously belong to multiple groups. Methods for uncovering overlapping communities in static networks have been recently proposed by Ahn et al. [24], and Evans and Lambiotte [25]. Temporal communities can also reveal overlapping communities in the aggregate network comprising all snapshots, since a node can participate in multiple communities over time. However, the definition of estrangement is easily generalized to the case where even within a snapshot nodes may belong to multiple overlapping communities. If the overlapping community membership of a node in a given snapshot is represented by a set of labels, we first define the *consort score* of an edge as the Jaccard similarity of label sets of the endpoint nodes. Estrangement is then defined as the sum over all edges, of the difference in consort score of an edge from time $t$ to $t-1$, divided by the number of edges. This definition clearly reduces to the definition in the main paper if the label sets are of size one, i.e., the communities are non-overlapping. If there are weights associated with the labels representing the overlapping community membership – where the weights represents the extent of participation of that node in the community – then we can use a generalization of Jaccard similarity such as Tanimoto similarity to compute the consort scores. Note that this generalized definition of estrangement is decomposable into node-local components as well, and thus – similar to our work for non-overlapping communities – methods for finding overlapping communities can be adapted to find overlapping temporal communities by constraining this generalized estrangement.

## 4.3 Proof that the HLPA update rule performs a greedy maximization of the Lagrangian

Here we show how Estrangement can be decomposed into node-local terms (similar to modularity) which enables the prescription of a label update rule (Eq. 6) which greedily maximizes the Lagrangian $L(\mathcal{P}, \lambda)$. Following Barber and Clark [12], we expand $Q$ and write $L(\mathcal{P}, \lambda)$

as:

$$L(\mathcal{P}, \lambda) = \frac{1}{2M}\left(\sum_u \sum_v (A_{uv} - \frac{1}{2M}k_u k_v + \lambda Z_{uv})\delta(l_u, l_v)\right) \tag{7}$$

Here we have taken advantage of the fact that the first term $\frac{\sum_{uv} Z_{uv}}{2M}$ in $E$ (Eq. 3) is independent of the partition and does not affect the optimization. To see the effect of a label update for a single node $x$, we separate terms of Eq. 7 into contributions from $x$ and those from all other nodes. Doing so yields:

$$
\begin{aligned}
L = \frac{1}{2M}&\left(\sum_{u \neq x}\sum_{v \neq x}(A_{uv} - \frac{1}{2M}k_u k_v + \lambda Z_{uv})\delta(l_u, l_v)\right) \\
&- \frac{1}{2M}\left(A_{xx} - \frac{1}{2M}k_x^2 + \lambda Z_{xx}\right) \\
&+ \frac{1}{2M}\left(2\sum_u (A_{ux} - \frac{1}{2M}k_u k_x + \lambda Z_{ux})\right)\delta(l_u, l_x)
\end{aligned} \tag{8}
$$

where in the interest of brevity, we have introduced the shortened notation $L$ to mean $L(\mathcal{P}, \lambda)$. The first two terms in $L$ ( R.H.S. of Eq. 8) are unaffected by the label update of node $x$, so we focus on the last term. Since our goal is to greedily optimize $L$ via label updates, we want the $x$ to update its label to one that will result in the maximal gain in $L$. Thus, the desired post-update label is:

$$
\begin{aligned}
l_x &= \arg\max_l \frac{1}{2M}\left(2\sum_u ((A_{ux} - \frac{1}{2M}k_u k_x + \lambda Z_{ux}))\right)\delta(l_u, l) \\
&= \arg\max_l \left(N_{xl} - \sum_{ux}\frac{1}{2M}k_u k_x \delta(l_u, l) + \sum_{ux}Z_{ux}\delta(l_u, l)\right)
\end{aligned} \tag{9}
$$

where we have used the fact that $\sum_u A_{ux}\delta(l_u, l)$ is simply the number of neighbors of $x$ with label $l$, which we denote by $N_{xl}$. The diagonal terms (i.e. terms with $u = x$) in the remaining sums of the above equation do not have any bearing on the maximization, and can be ignored. Then, using:

$$\sum_{u \neq x}\frac{1}{2M}k_u k_x \delta(l_u, l) = \frac{k_x K_l}{2M} - \frac{k_x^2}{2M}\delta(l_x, l)$$

and writing $\sum_{u \neq x} Z_{ux}\delta(l_u, l)$ as $O_{xl}$ (also, $K_l = \sum_u k_u\delta(l_u, l)$), we see that Eq. 9 reduces to:

$$l_x = \arg\max_l \left(N_{xl} - \frac{k_x K_l}{2M} + \frac{k_x^2}{2M}\delta(l_x, l) + \lambda O_{xl}\right),$$

This equation is identical to the HLPA update rule, Eq. 6. It therefore follows that the HLPA label update rule maximizes the gain in $L$.

## 4.4 Details on solving the dual problem

Having found a way to compute $g(\lambda)$ we can solve the dual problem and determine the value of $\lambda$ at which $g(\lambda)$ is minimized. The challenge here is that $g(\lambda)$ is not differentiable and moreover expensive to evaluate. We use Brent's method [21] which is often used to optimize non-differentiable scalar functions within a given interval. In our case, $g(\lambda)$ is the scalar function, and we minimize it within a suitably large range of $\lambda$. We use an implementation provided by python's scientific library, SciPy, in the form of scipy.optimize.fminbound(). For all experiments in this work, $\lambda_{min} = 0$ and $\lambda_{max} = 10$.

Furthermore, to mitigate issues due to the local nature of the algorithm and the degeneracy of the modularity landscape (and therefore the $L(\mathcal{P}, \lambda)$ landscape), we perform several independent runs of HLPA for a given $\lambda$ and pick the run which yields the highest value of $g(\lambda)$. We perform at least 10 runs of HLPA as we start Brent's method and increase the number of runs by 10 with every iteration that narrows the search interval for lambda. Near the optimum value of $\lambda$, we perform at least 150 runs to compute the Lagrange dual. Once we identify the value $\lambda = \lambda^*$ for which $g(\lambda)$ is minimized, the partition which yielded the highest $g(\lambda^*)$ over the many independent runs for $\lambda = \lambda^*$ is chosen as the optimal partition for the given snapshot. In practice, due to the degeneracy of the $L(P, \lambda)$ landscape for any $\lambda$, we have to go slightly above $\lambda^*$ to ensure that the optimal partition lies within the feasible region.

## 4.5 The Lagrangian computation is unaffected by the induce-graph operation

In HLPA, after the local update rule Eq. 6 has converged, a new graph is *induced* from the current graph and the partition that the local update rule has converged to. In this new graph, the communities of the converged partition play the role of the nodes, which we call "supernodes". Each supernode in the induced graph has a self-loop with a weight equal to twice the sum of weights of all links in the original graph contained in the community that forms the supernode. Similarly, a link between two supernodes has a weight equal to the sum of the weights of all edges connecting the two communities represented by the supernodes in the original graph.

Here we prove that the modularity $Q$ and the estrangement $E$ computed for a partition of an induced graph yields the same result as that computed when considering all nodes and edges within the supernodes of the induced graph. As a result, the Lagrangian computed for a partition on the induced graph is also identical to that computed when taking into account all nodes and edges contained within them. We use indices $u,v$ to refer to supernodes of an induced graph, and $i,j$ to refer to nodes of the original graph (contained within supernodes of the induced graph). Thus $u$ and $v$ also refer to community labels at the previous hierarchical level. For a given partition of the induced graph the modularity can be written as :

$$Q^{\text{induced}} = \sum_c \sum_{u,v \in c} \left( \frac{A_{uv}}{2M} - \frac{k_u k_v}{(2M)^2} \right)$$

where $c$ runs over indices of the different communities. We can write $A_{uv}$ as $\sum_{i \in u, j \in v} A_{ij}$. Similarly, $k_u = \sum_{i \in u} k_i$ and $k_v = \sum_{j \in v} k_j$ which gives:

$$Q^{\text{induced}} = \sum_c \sum_{u,v \in c} \left( \sum_{i \in u, j \in v} \frac{A_{ij}}{2M} - \sum_{i \in u, j \in v} \frac{k_i k_j}{(2M)^2} \right)$$

By transitivity of the community labels (i.e. $i \in u$ and $u \in c \implies i \in c$), we can therefore write:

$$Q^{\text{induced}} = \sum_c \sum_{i,j \in c} \left( \frac{A_{ij}}{2M} - \frac{k_i k_j}{(2M)^2} \right)$$

which is identical to the modularity of the same partition computed over nodes of the original graph. Similarly, the term which accounts for the contribution from estrangement due to a partition of the induced graph can be written as:

$$\sum_{u,v} \frac{Z_{uv}}{2M} \delta(l_u, l_v) = \sum_c \sum_{u,v \in c} \frac{Z_{uv}}{2M}$$

where, as before, $c$ runs over the indices of the different communities. We can write $Z_{uv} = \sum_{i \in u, j \in v} Z_{ij}$. Thus the estrangement term becomes:

$$\sum_c \sum_{u,v \in c} \sum_{i \in u, j \in v} \frac{Z_{ij}}{2M} = \sum_c \sum_{i,j \in c} \frac{Z_{ij}}{2M} = \sum_{ij} Z_{ij} \delta(l_i, l_j)$$

which is identical to the estrangement term computed on the original graph. It follows that for a given partition the Lagrangian is preserved when moving from the original graph to the induced graph.

## 4.6 Interpretation of estrangement confinement in terms of random walk temporal stability

Several quality functions for evaluating static community structure can be interpreted in terms of the properties of a random walk on some network construction. For example, in the case of modularity, [14, 2] have demonstrated that modularity can be reinterpreted as a random walk measure known as *stability*. Stability is based on the premise that the probability of a random walker to stay in the same community after a single step should be much larger than it would be on a fully randomized instantiation of the network. Specifically, stability compares the probability that a random walker at stationarity is in the same community after a step, with the analogous probability obtained for a random walker on a degree-sequence preserving null model of the network under consideration. The first term in this comparison is $\sum_{uv} \frac{A_{uv}}{2M} \delta(l_u, l_v)$ while the second term is $\sum_{uv} \frac{k_u k_v}{(2M)^2} \delta(l_u, l_v)$, thus yielding the following expression for stability:

$$S_{\text{structural}} = \frac{1}{2M} \sum_{uv} \left( A_{uv} - \frac{k_u k_v}{2M} \right) \delta(l_u, l_v)$$

Here, we refer to stability as defined in [14] as *structural* stability as it is a function solely of the structure of the network. Clearly, $S_{\text{structural}}$ is identical to modularity, $Q$.

The notion of stability can be extended to the case of temporal communities by incorporating an additional term which characterizes the *temporal stability* of a random walk. An intuitive explanation for this quantity can be given by considering the example of an unweighted network evolving from $t-1$ to $t$. Consider edges in $G_t$ that were also present in $G_{t-1}$, and additionally were intracommunity edges in $P_{t-1}$. We refer to such edges as *historical* edges.

Then, the temporal stability (of a partition $P_t$) compares the probability that a random walker in $G_t$ at stationarity, walks along a historical edge, and ends up in the same community, with the best case value of this probability. By best case value, we mean the value obtained in the maximally temporally stable case viz. the case where the chosen partition for $G_t$ makes every historical edge an intra-community edge. Thus, temporal stability measures the degree to which a random walker's environment remains invariant (i.e. it is in the same community) after a one step walk in $t$, given that it was invariant for a one step walk in $t-1$.

Formally, the temporal stability as defined above can be written as:

$$S_{\text{temporal}} = \sum_{uv} \frac{Z_{uv}}{2M} \delta(l_u, l_v) - \sum_{uv} \frac{Z_{uv}}{2M}$$

where $l_u, l_v$ are defined by the partition $P_t$ under consideration, and $Z_{uv}$ is as defined in Methods. Here, the second term is the one obtained for the maximally temporally stable case. As is clear from Eq. 3, $S_{\text{temporal}} = -E$. Thus, $S_{\text{temporal}} \leq 0$.

It follows that the constrained optimization problem to be solved for finding temporal communities, Eq. 1 is equivalent to the problem of maximizing structural stability, $S_{\text{structural}}$, while constraining temporal stability, $S_{\text{temporal}}$, to be greater than or equal to $-\delta$ (where $\delta$ is non-negative). The analogy of temporal stability with estrangement can be generalized to the case of weighted networks as well.